\pdfoutput=1
\documentclass[onecolumn,amsmath,amssymb,showpacs]{revtex4-1}
\usepackage{graphicx,color,xcolor,hyperref,amssymb,textcomp,amsmath,bm}
\usepackage[normalem]{ulem}
\allowdisplaybreaks[1]

\newcommand{\eref}[1]{Eq.~(\ref{#1})}%
\newcommand{\fref}[1]{Fig.~\ref{#1}} %
\usepackage{mwe}
\usepackage{mathtools}

\def\bea{\begin{eqnarray}}
\def\eea{\end{eqnarray}}

\DeclareMathOperator{\Var}{Var}

\begin{document}

\title{Thermodynamic uncertainty relation for systems with unidirectional transitions}

\author{Arnab Pal$^{1}$}
\email{arnabpal@mail.tau.ac.il}

\author{Shlomi Reuveni$^{1}$}
\email{shlomire@tauex.tau.ac.il}

\author{Saar Rahav$^{2}$}
\email{rahavs@ch.technion.ac.il}

\affiliation{\noindent \textit{$^{1}$School of Chemistry, The Center for Physics and Chemistry of Living Systems, \& The Mark Ratner Institute for Single Molecule Chemistry, \& The Raymond and Beverly Sackler Center for Computational Molecular and Materials Science, Tel Aviv University, Tel Aviv 6997801, Israel}}

\affiliation{\noindent \textit{$^{2}$Schulich Faculty of Chemistry, Technion-Israel Institute of Technology, Haifa 32000, Israel}}

\date{\today}
\begin{abstract}
We derive a thermodynamic uncertainty relation (TUR) for systems with unidirectional transitions.  The uncertainty relation involves a mixture of thermodynamic and dynamic terms. Namely, the entropy production from bidirectional transitions, and the net flux of unidirectional transitions. The derivation does not assume a steady-state, and the results apply equally well to transient processes with arbitrary initial conditions. As every bidirectional transition can also be seen as a pair of separate unidirectional ones, our approach is equipped with an inherent degree of freedom. Thus, for any given system, an ensemble of valid TURs can be derived. However, we find that choosing a representation that best matches the system’s dynamics over the observation time will yield a TUR with a tighter bound on fluctuations. More precisely, we show a bidirectional representation should be replaced by a unidirectional one when the entropy production associated with the transitions between two states is larger than the sum of the net fluxes between them. Thus, in addition to offering TURs for systems where such relations were previously unavailable, the results presented herein also provide a systematic method to improve TUR bounds via physically motivated replacement of bidirectional transitions with pairs of unidirectional transitions. The power of our approach and its implementation are
demonstrated on a model for random walk with stochastic resetting 
and on the Michaelis-Menten model of enzymatic catalysis.
\end{abstract}
\maketitle

\section{Introduction}
\label{sec:intro}
\noindent
The last three decades have seen significant progress in our understanding of out-of-equilibrium systems and processes. The celebrated fluctuation theorem replaces the inequality of the second law by an equality that connects the ratio of probabilities of symmetry related realizations to thermodynamic quantities  \cite{Evans1993,Gallavotti1995,Kurchan1998,Lebowitz1999,Seifert2005,Jarzynski1997,Crooks1998}.
This important result have shown the usefulness of assigning thermodynamical interpretation to single realizations of an out-of-equilibrium process. A theoretical framework, termed stochastic thermodynamics, was built around this idea \cite{Sekimoto-book,Seifert2012}. Stochastic thermodynamics is well suited to describe single molecule experiments of molecular motors and machines which operate while being immersed in liquid environments
\cite{Book}.

One of the intriguing results in the field is the thermodynamic uncertainty relation (TUR) \cite{Barato2015,Gingrich2016,Horowitz2019}. The TUR is a bound involving
the mean value of a fluctuating current $J_{\mathcal{T}}$, its variance Var$[J_{\mathcal{T}}]$, and the average entropy production $\Sigma_{\mathcal{T}}$ accumulated upto time $\mathcal{T}$. In steady state, it takes the form 
\begin{equation}
\frac{\Var \left[J_{\mathcal{T}}\right]}{\langle J_{\mathcal{T}} \rangle^2} \ge \frac{2}{\Sigma_{\mathcal{T}}}~,
\end{equation} 
in units where the Boltzmann's constant is set to $k_\text{B}=1$.
Loosely speaking the TUR reveals that beyond a certain threshold, 
variance reduction, or increased precision, can only be obtained at the expense of increased dissipation. The TUR can be used to obtain bounds on the entropy production of a system without the need to know specific details about its structure \cite{Pietzonka2016}.
The TUR was first suggested by Barato and Seifert \cite{Barato2015}, based on the study of several models, and was then derived using large deviation theory \cite{Gingrich2016}. The simplicity, appealing physical interpretation, and generality of the TUR have led to many extensions and related results \cite{Polettini2016,Proesmans2017,Horowitz-2017,Pietzonka2017,Gingrich2017, Garrahan2017,Barato2018,Nardini2018,Koyuk2018,Dechant2018,Terlizzi2018,Carollo2019,Potts2019,Lee2019,Vu2020,Falasco2020,quantum1,quantum2,EP-short,TUR-expt}.
Two mathematical approaches were used to derive the TUR and its generalizations. The original proof was based on the large deviations formalism \cite{Gingrich2016}.
An alternative approach, based on the Cram\'er-Rao inequality have been used to re-derive the TUR and related results \cite{Dechant2018b,Hasegawa2019,Dechant2020}. This information theoretic approach will also be used below.

To date, work on the TUR was mainly focused on systems that are in steady-state. The inability to describe fully time dependent processes was a major limitation of the theory. This serious gap in the theory was only closed very recently. In \cite{Liu2019}, Liu {\em et. al.} derived a TUR that is valid for systems with arbitrary initial states, and is thus applicable for finite-time relaxation processes. While this manuscript was being written, Koyuk and Seifert have presented a TUR that applies for processes with time dependent rates \cite{Koyuk2020}. These two works significantly extend the applicability of TURs. 

There is still a class of systems for which the TURs do not apply, namely systems with unidirectional transitions.  Here, a unidirectional transition is one which has a  finite
rate $R_{ij}>0$, while its reversed counterpart is forbidden, namely $R_{ji}=0$. Systems with unidirectional transitions are less studied since
much of stochastic thermodynamics is built upon local detailed balance, which can hold only for bidirectional transitions. 
Thus, it is not surprising that only a few papers were devoted to the stochastic thermodynamics of systems with unidirectional transitions  \cite{BA2011, Zeraati2011,Murashita2014, Rahav2014,Baiesi2015,Fuchs2016,Pal2017,Busiello2020,workflucPRL2020}.

Nevertheless, there are many instances where one is interested in physically relevant models that
include unidirectional transitions. These may appear as an idealization of a process 
whose inverse rarely occurs
on a relevant time-scale, or because they are meant to represent externally controlled events such as resetting (to be described below). As relevant examples consider the total asymmetric simple exclusion process (TASEP) \cite{Derrida},  driven dissipative systems e.g., the inelastic Lorentz gas \cite{dds}, directed percolation in liquid crystals \cite{percolation}, and the decay of an atom via spontaneous emission \cite{Scullybook}. Such irreversible transitions also occur in cytoneme based morphogenesis \cite{Bressloff1}, motor driven intracellular transport \cite{ Bressloff2}, backtracking recovery in RNA polymerization \cite{Edgar-RNA}, and in models of population dynamics \cite{population} and queuing \cite{queue} where irreversible transitions are manifested as catastrophes.
Unidirectional transitions are also used to model chemical enzymatic reactions, in particular, the catalytic process \cite{enzyme2,enzyme3,enzyme4}. Quite ubiquitously, they also appear in discrete models of first passage problems \cite{Redner}. 
 
A particular set of unidirectional transitions that has recently drawn considerable attention arises in systems with resetting. There, upon resetting,
the system is returned to a predetermined state. Stochastic resetting was
studied in connection with
random search processes. Interestingly, it was found that the addition of resetting can reduce the mean time taken to
complete the search, due to elimination of realizations with extremely long search completion times. This seminal result has led to an extensive research effort focused on the properties of stochastic resetting systems
\cite{EM11a,SR-review,KMSS14,EM14,Pal14,SabhapanditTouchette15,MSS15,time-dep,Eule2016,RoldanGupta2017,Shamik16,SR2016,PalShlomi2016,Branching,Landau,interval}. In addition, resetting was recently realized experimentally \cite{expt1,expt2}. We refer to \cite{SR-review} for an extensive review of the subject.
In stark contrast, to date only a few papers were devoted to the stochastic thermodynamics of resetting.
Fuchs {\em et. al.} used stochastic thermodynamics to give a consistent thermodynamic interpretation of
the resetting processes \cite{Fuchs2016} and derived the first and second law for them. Stochastic resetting systems were also shown to satisfy integral fluctuation theorems in \cite{Pal2017}. Recently, universality of work fluctuations followed up by the validation of the Jarzynski equality was studied for a stochastic resetting system  \cite{workflucPRL2020}. Yet, a TUR for systems with stochastic resetting --- and more generally for systems with \textit{unidirectional transitions} --- is still missing.

In this work we present a TUR for systems with unidirectional transitions. Our derivation is motivated by, and follows closely, the approach of Liu {\em et. al.} \cite{Liu2019}, but nevertheless extends it in \textit{two important aspects}. First, we modify the original derivation to apply to unidirectional transitions. In addition, we also allow for non current-like observables, such as the time spent in different states. Interestingly, the TUR we obtain includes a mixture of thermodynamic and dynamic contributions. Specifically, the entropy production term in the familiar TUR is replaced with a linear combination of the entropy production due to bidirectional transitions and the total flux (or activity) of the unidirectional transitions. The TUR we derive
can be applied to bound the fluctuations of a diverse set of observables in various different setups. To illustrate the versatility of the formalism, we will demonstrate its power in and out of steady-state. We will also show how the freedom to treat bidirectional transitions as pairs of separate unidirectional transitions gives rise to a systematic, and physically motivated, method to improve bounds on fluctuations and make them tighter.

Our paper is organized as follows. In Sec. \ref{sec:model} we present the models that will be studied, namely Markovian jump processes with a combination of unidirectional and bidirectional transitions.
 In Sec. \ref{sec:derivation} we generalize the derivation of Liu. {\em et. al.} \cite{Liu2019} to systems with unidirectional transitions.
In Sec. \ref{sec:application} we present two applications of the TUR. The first is for the number of resetting events in a stochastic resetting system that is in a steady state. The second is for the 
probability to complete an enzymatic catalytic process by a certain
finite time. We conclude in Sec. \ref{sec:conc}.

\section{Model and setup}
\label{sec:model}
\noindent
We consider systems whose dynamics can be modeled as a Markovian jump process on a network with a finite number of states, $N_s$. The physical properties of the system are determined by
the connectivity of the network
and the dependence of the transition rates on physical parameters. Let us denote the rate of the transition from state $j$ to $i$ by $K^{(\alpha)}_{ij} (t)$. Here, $\alpha$ is an index that is used to distinguish between several
physically different transitions that occur between the same two states, e.g., due to coupling to different temperatures or particle reservoirs. 
The principle of micro-reversibility states
that if $K^{(\alpha)}_{ij}>0$ then also $K_{ji}^{(\alpha)}>0$.
In stochastic thermodynamics, it is also common to demand local detailed balance, for instance %,
\begin{equation}
 \frac{K^{(\alpha)}_{ij}}{K_{ji}^{(\alpha)}}= \exp \left[ \beta_\alpha \left( E_j - E_i\right)\right],
\label{eq:ldb}
\end{equation}
where $E_i$ is the energy of the $i$-th state. The principle of local detailed balance is based on the assumption that the transition is coupled to an environment that is in equilibrium with an inverse 
temperature $\beta_\alpha$. The condition (\ref{eq:ldb}) is needed for the model to be thermodynamically consistent.  \eref{eq:ldb} is just an example for local detailed balance in a process where energy is exchanged. It should be modified if the transition involves an exchange of particles, or an externally applied non-conservative force. A thorough discussion of local detailed balance can be found in the review of stochastic thermodynamics by Seifert \cite{Seifert2012}. In what follows we will use the term bidirectional transitions to refer to transitions like the ones described above.

In addition to bidirectional transitions, we further allow %the systems to have
unidirectional transitions between states. These are denoted by rates $R^{(\gamma)}_{ij}>0$,
with a reversed transition whose rate vanishes, $R^{(\gamma)}_{ji}=0$. Here, $\gamma$ makes a distinction between unidirectional transitions, that occur between the same two states but are 
of different physical origin. 
Clearly, such transitions violate microreversibility and local detailed balance.
We intentionally use different symbols for bidirectional and unidirectional transitions, since distinguishing them will help clarify many of the subsequent calculations.  We note that while the distinction between unidirectional and bidirectional transitions is meant to represent properties of the model of interest (such as processes which almost never occur), there is nothing that prevents
one from formally viewing a bidirectional transition as a pair of unidirectional ones. This freedom will
be used later to clarify some aspects of the approximations that allow to treat transitions as unidirectional.

\begin{figure}[h]
\includegraphics[scale=0.5]{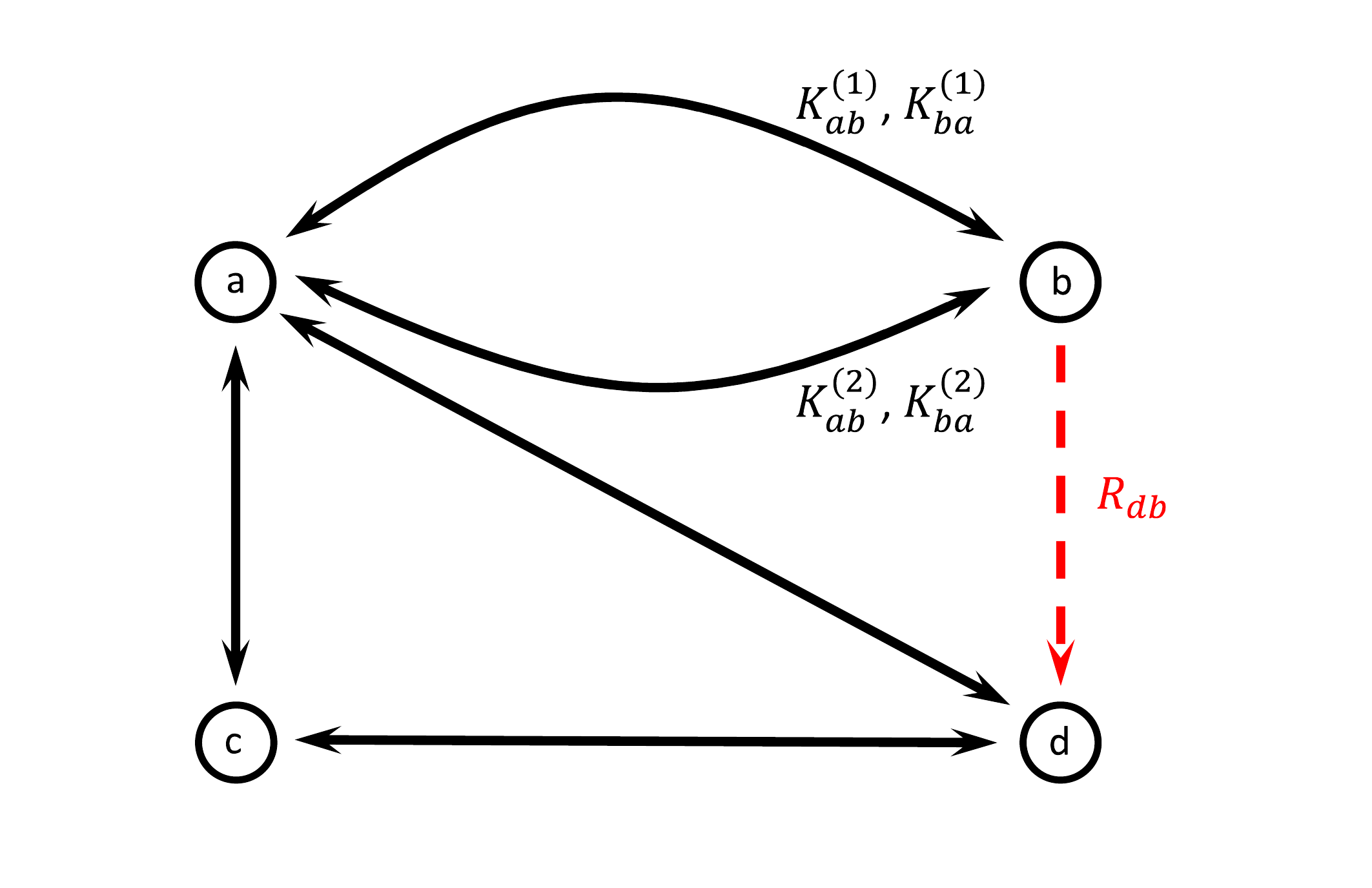}
\caption{An example for a graph representing a Markovian jump process. The model has four states, with five bidirectional transitions connecting them. These transitions (e.g., $K_{ab}^{(1)}, K_{ba}^{(1)}$) are denoted by two-sided arrows. Note that states $a$ and $b$ are connected by two distinct transitions with different rates (e.g. $K_{ab}^{(1)}$ and $K_{ab}^{(2)}$ ). These are assumed to have different physical origin.
The transition from state $b$ to $d$ (denoted by $R_{db}$) is unidirectional, and is represented by a one-sided, red, dashed, arrow.
\label{fig:graph}}
\end{figure}

It is often convenient to depict a Markovian jump process as a graph, as is done in Fig. \ref{fig:graph}. The nodes of the graph correspond to the states of the model, whereas the links denote the allowed transitions.
The net rate of transitions from state $j$ to state $i$ is
\begin{equation}
\Gamma_{ij}(t)=\sum_{\alpha}~ K_{ij}^{(\alpha)}(t)+\sum_{\gamma}~ R_{ij}^{(\gamma)}(t).
\label{eq:defgnd}
\end{equation}
It is also useful to consider the escape rate out of state $j$
\begin{equation}
\lambda_j(t)=-\Gamma_{jj}(t)=\sum_{\substack{i=1 \\ i \ne j}}^{N_s}~\Gamma_{ij}(t).
\label{eq:defgd}
\end{equation}
The probability to find the system in each state evolves according to a master equation
\begin{equation}
\frac{d \mathbb{P}}{dt}=\bm{\Gamma} \mathbb{P},
\label{ME}
\end{equation}
where $\mathbb{P}(t)$ denotes a vector containing the probabilities to find the system in
the different states at time $t$, and the matrix elements of $\bm{\Gamma}$ are given in Eqs. (\ref{eq:defgnd}) and (\ref{eq:defgd}).

If we follow the state of the system as a function of time we will observe a realization, or a history, of the process.
Each history is characterized by a list of states the system was in, and the transitions the system made to pass between them. An example for such a
history is $\omega=(i_0,t_0=0;i_1,t_1, \xi_1;i_2,t_2, \xi_2; \cdots;i_n,t_n, \xi_n \leq \mathcal{T})$,  
where the transition from $i_{l-1}$  to $i_l$ happened at time $t_l$. Here $\xi$ is an index that identifies the transition that has taken place. It will point out to a bidirectional transition if it matches one of the $\alpha$'s, and to a unidirectional one if its value matches one of the $\gamma$'s that are allowed for this transition. Such histories (possibly coarse-grained) are observed in single molecule experiments on certain molecular motors and machines.
To understand such stochastic processes we should consider all possible histories with a given final time, $\mathcal{T}$, and their probabilities. 
Note that this family of histories include ones with different numbers of transitions, $n$.
Let us say that we are interested in a physical quantity $\mathcal{F}\left[ \omega \right]$
that can be calculated for each history $\omega$. With the help of the probability density $\mathcal{P} \left[ \omega \right]$ of histories we can discuss its mean value
$\langle \mathcal{F} \rangle$ and its fluctuations. In the following we will need to consider both time-independent and time-dependent processes. We start by examining the simpler time-independent case.

For systems with time independent transition rates, we can characterize each history by the number of times each bidirectional transition was made, which is given by
\begin{equation}
{n}_{ij}^{(\alpha)} \left[  \omega \right]=\sum_{l=1}^n~\delta_{i_{l-1},j}~ \delta_{i_{l},i} ~ \delta_{\xi_{l},\alpha} ,
\end{equation}
and equivalently for unidirectional transitions. 
Another relevant quantity is the time spent in each state (i.e., the residence time) during such a history
\begin{equation}
\tau_i \left[ \omega \right]=\sum_{l=1}^n~\delta_{i_l,i} \left( t_{l+1}-t_l \right).
\label{eq:deftau}
\end{equation}
Here we use the convention that $t_{n+1}= \mathcal{T}$, or equivalently $\tau_n  \left[ \omega \right] \equiv \mathcal{T}-t_n$, to write Eq. (\ref{eq:deftau}) in a more compact form. This should not be taken to mean that there is a transition at $t_{n+1}= \mathcal{T}$. The probability density of the history $\omega$ is then given by \cite{Seifert2012,Pal2017}
\begin{equation}
\mathcal{P}[\omega] \equiv P_{i_0}(0)~ \exp \left( -\sum_{i=1}^{N_s}~\lambda_i~ \tau_i \left[ \omega \right]~ \right) 
\exp \left( \sum_{\substack{i,j,\alpha \\ i \ne j}} n_{ij}^{(\alpha)} \left[  \omega \right] \ln {K}_{ij}^{(\alpha)} \right)
\exp \left( \sum_{\substack{i,j,\gamma \\ i \ne j}}n_{ij}^{(\gamma )} \left[  \omega \right] \ln {R}_{ij}^{(\gamma)} \right),
\label{eq:path-measure}
\end{equation}
where $P_{i_0}(0)$ is the initial condition.
The sum $\sum_{\substack{i,j \\ i \ne j}} X_{ij} = \sum_{j=1}^{N_s} \sum_{\substack{i=1 \\ i\ne j}}^{N_s} X_{ij}$ is a compact way of writing the sum over all transitions (ordered pairs of states).
In what follows we adopt a notation in which the top subscript in a summation symbol denotes the variables being summed, while the bottom subscript gives additional restrictions, such as $i \ne j$ or $i>j$. 

For processes with time independent rates, the physical quantities we will be interested in, are given  
by functionals of the form
\begin{equation}
\mathcal{F}[\omega] \equiv \underbrace{ \sum_{i=1}^{N_s}~q_i \tau_i \left[ \omega \right]}_{\text{residence time}} + \underbrace{ \sum_{\substack{i, j,\alpha \\ i \ne j}}d_{ij}^{(\alpha)} n_{ij}^{(\alpha)}  \left[ \omega \right] }_{\text{bidirectional jumps}} + \underbrace{ \sum_{\substack{i , j, \gamma \\ i \ne j }}c_{ij}^{(\gamma)}n_{ij}^{(\gamma)}  \left[ \omega \right] }_{\text{unidirectional jumps}},
\label{eq:functional}
\end{equation}
where $q_i$, $d_{ij}^{(\alpha)}$, and $c_{ij}^{(\gamma)}$ are parameters that can be chosen so that $\mathcal{F}$ can describe different physical quantities.
The first term in Eq. (\ref{eq:functional}) measures quantities related to residence times. The second term quantifies the contribution of bidirectional transitions. Here often one demands that
$d_{ij}^{(\alpha)}$ is antisymmetric, namely that $d_{ij}^{(\alpha)}=-d_{ji}^{(\alpha)}$. The reason is that many physical quantities, including various currents and entropy production are obtained from antisymmetric $d_{ij}$. The derivation of the TUR below utilizes
this requirement in order to make a connection with the entropy production of bidirectional
transitions.
The last term in Eq. (\ref{eq:functional}) gives the contribution from unidirectional transitions. Consequently,  $c_{ij}^{(\gamma)}$ need
not be antisymmetric.

For time dependent processes the form of the probability density of histories and the functionals are more cumbersome
as they depend on the values of the rates over the entire history course. In this case, it is convenient to define
\begin{equation}
\chi_i (t) = \delta_{\omega(t),i},
\end{equation}
which is an indicator function that attains 
the value $1$ if the system is in state $i$ at time $t$, and $0$ otherwise. We also use
\begin{equation}
\Dot{n}^{(\alpha)}_{ij} (t) =\sum_{l=1}^n ~\delta_{i_{l-1},j}~ \delta_{i_{l},i} ~ \delta_{\xi_{l},\alpha} \delta \left(t-t_l \right),
\end{equation}
which is a sum of Dirac delta functions at the times that match the bidirectional transition $j \rightarrow i$, via the $\alpha$ channel; and equivalently for unidirectional transitions via the $\gamma$ channel. Using these, the probability density of a history $\omega$ in a time dependent process can be written as \cite{Seifert2012,Pal2017}
\begin{equation}
\mathcal{P}[\omega(t)] \equiv P_{i_0}(0)~ \exp \left[ -\int_0^{\mathcal{T}} dt \left( \sum_{i=1}^{N_s}~\lambda_i(t)~ \chi_i (t)-
 \sum_{\substack{i , j, \alpha \\ i \ne j}} \dot{n}_{ij}^{(\alpha)} (t) \ln {K}_{ij}^{(\alpha)}(t) -  \sum_{\substack{i, j, \gamma \\ i \ne j}} \dot{n}_{ij}^{(\gamma)} (t) \ln {R}_{ij}^{(\gamma)}(t) \right) \right] .
\label{eq:path-measure-time-dep}
\end{equation}

The probabilities of histories are normalized such 
that $\sum_\omega \mathcal{P} [\omega]=1$, where the sum over histories should be understood as a sum over the number of transitions and 
integration over all the intermediate times.
We note that the ensemble average over the histories of $\chi_i (t)$ is the probability to be at state $i$ at time $t$,
\begin{equation}
\langle \chi_i (t)  \rangle=\sum_{\omega(t)}\mathcal{P}[\omega(t)] \delta_{\omega(t),i}  =P_i(t).
\label{exp1}
\end{equation}
Similarly, we have
\begin{equation}
\langle \Dot{n}^{(\alpha)}_{ij} (t) \rangle = {K}_{ij}^{(\alpha)}(t) P_j (t),
\label{exp2}
\end{equation}
which is the flux through channel $\alpha$ of bidirectional
$j \rightarrow i$ transitions
at time $t$; and equivalently for unidirectional transitions  via the channel $\gamma$.
For time-dependent processes, one may consider more general functionals, of the following form 
\begin{equation}
\mathcal{F}[\omega] \equiv \int_0^{\mathcal{T}} dt \left( \sum_{i=1}^{N_s}~q_i (t) \chi_i (t) + \sum_{\substack{i , j, \alpha \\ i \ne j}} d_{ij}^{(\alpha)}(t) \Dot{n}_{ij}^{(\alpha)}(t)  +\sum_{\substack{i , j, \gamma \\ i \ne j}} c_{ij}^{(\gamma)} (t) \Dot{n}_{ij}^{(\gamma)}(t)  \right).
\label{eq:tdfunctional}
\end{equation}
%\begin{equation}
%\mathcal{F}[\omega] \equiv \int_0^{\mathcal{T}} dt \left( \sum_{i=1}^{N_s}~q_i (t) \chi_i (t) + \sum_{(i \neq j), \alpha}\Dot{n}_{ij}^{(\alpha)} (t)  d_{ij}^{(\alpha)}(t) +\sum_{(i \neq j), \gamma}\Dot{n}_{ij}^{(\gamma)}(t)  c_{ij}^{(\gamma)} (t)\right).
%\label{eq:tdfunctional}
%\end{equation}
\noindent This expression allows to consider time dependent weights $q_i (t)$ and counting fields $d_{ij}(t)$, $c_{ij}(t)$.
In what follows, we will
mostly be interested in systems with time-independent rates and physical quantities that are described by the time-independent functional in \eref{eq:functional}.
Note, however, that the derivation of the TUR requires us to also consider time-dependent extensions of the dynamics, and we will therefore need to use Eqs. (\ref{eq:path-measure-time-dep}) and (\ref{eq:tdfunctional}) as well.

Consider the mean value of the functional, $F(\mathcal{T})=\langle\mathcal{F}[\omega]  \rangle$. Substitution of Eqs. (\ref{exp1}) and (\ref{exp2}) in Eq. (\ref{eq:tdfunctional}) gives
\begin{equation}
F(\mathcal{T})=\int_0^{\mathcal{T}} dt~ \left[ \sum_{i=1}^{N_s}q_i (t) P_i(t)+ \sum_{\substack{j , i, \alpha \\ j \ne i}}
  d_{ij}^{(\alpha)} (t) ~{K}_{ij}^{(\alpha)} (t) P_j(t)+\sum_{\substack{j , i, \gamma \\ j \ne i}}
 c_{ij}^{(\gamma)} (t)~{R}_{ij}^{(\gamma)} (t)   P_j(t)\right],
\label{fT}
\end{equation}
and equivalently
\begin{equation}
\frac{d F}{dt} = \sum_{i=1}^{N_s}q_i(t) P_i(t)+\sum_{\substack{j , i, \alpha \\ j\ne i}}  d_{ij}^{(\alpha)} (t) K_{ij}^{(\alpha)} (t) P_j (t) + \sum_{\substack{j , i, \gamma \\ j \ne i}}  c_{ij}^{(\gamma)} (t) R_{ij}^{(\gamma)} (t) P_j (t).
\label{eq:dFdt}
\end{equation}
\eref{fT} will be the starting point for the derivation of the uncertainty relation. It is useful since it does not require enumerating all the histories of a process. It is therefore comparatively easy way of computing the mean value of a functional, as it only
requires the solution of the master equation, and the calculation of a simple integral.
An alternative derivation of Eq. (\ref{eq:dFdt}) is presented in Appendix \ref{sec:mean}.

\section{TUR with unidirectional transitions}
\label{sec:derivation}
\noindent
In this section we derive a TUR for jump processes with unidirectional transitions. Our derivation
extends the one presented in Ref. \cite{Liu2019} to systems with {\em unidirectional transitions }, and is similarly based on the
Cram\'er-Rao inequality.  Consider a parameter dependent extension of the dynamics that is obtained by allowing the transition rates to
depend on a parameter $\theta$. The rates $K_{ij,\theta}^{(\alpha)} (t)$ and $R_{ij,\theta}^{(\gamma)} (t)$
are assumed to depend smoothly on $\theta$, and reduce to the physical dynamics at $\theta=0$. The physical, or equivalently the original, dynamics ($\theta=0$) that we consider will have time independent rates. However, the system need not be at steady state since we will not make any demands regarding the initial conditions.

Although the initial condition, $P_{i} (0)$, is $\theta$-independent, the modified dynamics has a $\theta$-dependent probability distribution of histories, given by Eq. (\ref{eq:path-measure-time-dep}) with the
$\theta$-dependent rates. Considering all the possible histories between $t=0$ and $t=\mathcal{T}$, one can view $\mathcal{F}[\omega]$ as a random variable, with probability density $\mathcal{P}_\theta [\omega]$. The mean $F_\theta \left(\mathcal{\tau} \right)=\langle \mathcal{F} [\omega] \rangle$
and the variance of this random variable satisfies the generalized Cram\'er-Rao inequality \cite{CR1,CR2,CR3}
\begin{equation}
\text{Var}_{\omega} \big[ \mathcal{F}_\theta \left[\omega \right]  \big]  \geq \frac{\left[ \frac{\partial F_\theta (\mathcal{T})}{\partial \theta}  \right]^2}{\mathcal{I}(\theta)},
\label{eq:defgcr}
\end{equation}
where the Fisher Information is given by  \cite{CR1,CR2,CR3}
\begin{equation}
\mathcal{I}(\theta)=-\left \langle \frac{\partial^2}{\partial \theta^2} \ln \mathcal{P}_{\theta}(\omega) \right \rangle_{\omega}.
\label{eq:defFI}
\end{equation}
When $\theta=0$, the variance in \eref{eq:defgcr} is the variance of the functional $\mathcal{F} [\omega]$ in the physical (original) dynamics.
However, the terms on the right hand side of \eref{eq:defgcr} generally do not offer meaningful physical interpretation. The derivation of the TUR consists of identifying a correct parametrization of the transition rates that results in a physically meaningful bound. This will be done in the following.

The terms on the right
hand side of \eref{eq:defgcr} can be computed using the technique described in Sec. \ref{sec:model} and Appendix \ref{sec:mean}, albeit for
the process with the $\theta$-dependent rates. Using \eref{fT} for $F_\theta (\mathcal{T})$, we get 
\begin{equation}
F_{\theta}(\mathcal{T})=\int_0^{\mathcal{T}} dt~ \left( \sum_{i=1}^{N_s} q_i (t) P_{i,\theta}(t)+\sum_{\substack{i , j, \alpha \\ i \ne j}}
d_{ij}^{(\alpha)} (t) ~K_{ij,\theta}^{(\alpha)} (t)   P_{j,\theta}(t)+\sum_{\substack{i , j, \gamma \\ i \ne j}} c_{ij}^{(\gamma)} (t)
R_{ij,\theta}^{(\gamma)} (t)  P_{j,\theta}(t)  \right).
\label{F-theta}
\end{equation}
Taking a derivative with respect to $\theta$ and then taking the limit $\theta \rightarrow 0$ gives
\begin{multline}
\left. \frac{\partial F_{\theta}(\mathcal{T})}{\partial\theta} \right|_{\theta=0}= \int_0^{\mathcal{T}} dt~ \left(  \sum_{i=1}^{N_s} q_i (t) \left.\frac{\partial P_{i,\theta}(t)}{\partial \theta} \right|_{\theta=0}
+\sum_{\substack{i, j, \alpha \\ i \ne j}}
d_{ij}^{(\alpha)} (t) \left[ \frac{\partial P_{j,\theta}(t)}{\partial\theta} K_{ij}^{(\alpha)} (t) + \frac{\partial K_{ij,\theta}^{(\alpha)}(t)}{\partial \theta} P_{j}(t) \right]_{\theta=0} \right. \\
+\left. \sum_{\substack{i , j, \gamma \\ i \ne j}}
c_{ij}^{(\gamma)} (t) \left[  \frac{\partial P_{j,\theta}(t)}{\partial \theta} R_{ij}^{(\gamma)} (t) + \frac{\partial R_{ij,\theta}^{(\gamma)} (t)}{\partial \theta} P_{j}(t) \right]_{\theta=0} \right) .
\label{eq:dftheta}
\end{multline}
The Fisher information can be obtained by substituting the $\theta$-dependent version of Eq. (\ref{eq:path-measure-time-dep}) into
Eq. (\ref{eq:defFI}). This reveals that $\mathcal{I} (\theta)$ is the mean value of a functional of the form (\ref{F-theta}) with
\bea
q_i (t) &=& \frac{\partial^2 \lambda_{i,\theta}(t)}{\partial \theta^2}, \\
d_{ij}^{(\alpha)} (t) &=& -\frac{\partial^2 \ln K_{ij,\theta}^{(\alpha)} (t)}{\partial \theta^2},\\
c_{ij}^{(\gamma)} (t) &=& -\frac{\partial^2 \ln {R}_{ij,\theta}^{(\gamma)} (t)}{\partial \theta^2}.
\eea
Note that we used the fact that the initial probability does not depend on the parameter $\theta$. As a result, the Fisher information can be recast as
\begin{equation}
\mathcal{I}(\theta)=\int_0^{\mathcal{T}} dt~ \left( \sum_{i=1}^{N_s}  \frac{\partial^2 \lambda_{i,\theta} (t)}{\partial \theta^2} P_{i,\theta}(t)- \sum_{\substack{i , j, \alpha \\ i \ne j}}
~\frac{\partial^2 \ln {K}_{ij,\theta}^{(\alpha)} (t)}{\partial \theta^2}~ {K}_{ij,\theta}^{(\alpha)}(t)~  P_{j,\theta}(t) - \sum_{\substack{i , j, \gamma \\ i \ne j}}
~\frac{\partial^2 \ln {R}_{ij,\theta}^{(\gamma)}(t)}{\partial \theta^2} ~{R}_{ij,\theta}^{(\gamma)} (t) ~  P_{j,\theta}(t) \right). ~~
  \label{FI-T}
\end{equation}
The expression for $\mathcal{I} (\theta)$ can be simplified with the help of Eqs. (\ref{eq:defgnd}) and (\ref{eq:defgd}). After a bit of algebra, and taking the limit $\theta \rightarrow 0$, we obtain
\begin{equation}
\mathcal{I}(0)=
\int_0^{\mathcal{T}} dt~ \left( \sum_{\substack{i , j, \alpha \\ i \ne j}}
  P_{j}(t)~{K}_{ij}^{(\alpha)} (t)\left[ \frac{\partial \ln{K}_{ij,\theta}^{(\alpha)} (t) }{\partial \theta}  \right]^2_{\theta=0} 
  + \sum_{\substack{i , j, \gamma \\ i \ne j}}
  P_{j}(t)~{R}_{ij}^{(\gamma)}(t)\left[ \frac{\partial \ln{R}_{ij,\theta}^{(\gamma)} (t)}{\partial \theta}  \right]^2_{\theta=0}   \right).
\label{eq:FI-rates}
\end{equation}

To evaluate the first derivatives in Eqs. (\ref{eq:dftheta}) and (\ref{eq:FI-rates}) at $\theta=0$, we employ a small $\theta$-expansion. 
We first note that the probabilities $P_{i,\theta}$ in Eqs. (\ref{eq:dftheta}) and (\ref{eq:FI-rates}) are the solutions of the master equation
\bea
\dot{\mathbb{P}}_{\theta}(t)=\bm{\Gamma}_{\theta}(t)\mathbb{P}_{\theta}(t),
\label{master_theta}
\eea
where $\bm{\Gamma}_{\theta}(t)$ is the rate matrix built from the $\theta$-dependent rates. Recalling \eref{ME} where $\Gamma(t)$ is the rate matrix for the original dynamics, we expand the probability and the rate matrix in \eref{master_theta} for small $\theta$
\bea
\bm{\Gamma}_{\theta}(t) &= & \bm{\Gamma}+\theta \bm{\Gamma}_1(t)+\mathcal{O}(\theta^2), \\
\mathbb{P}_{\theta}(t)& =& \mathbb{P}(t)+\theta \mathbb{P}_1(t)+\mathcal{O}(\theta^2).
\eea
We demand that the parametrization satisfies 
\begin{equation}
  \bm{\Gamma}_1(t) \mathbb{P}(t)=\bm{\Gamma} \mathbb{P}(t)=\dot{\mathbb{P}}(t).
  \label{eq:condition1}
\end{equation}
The condition in \eref{eq:condition1} allows to calculate $ \mathbb{P}_1(t)$ with the  help of a simple linear response calculation
\begin{equation}
 \mathbb{P}_1(t) = \int_0^t dt' e^{\bm{\Gamma} (t-t')}  \bm{\Gamma}_1(t')  e^{\bm{\Gamma} t'} \mathbb{P}(0) = t\dot{\mathbb{P}} (t).
\end{equation}
As a result, we have $\mathbb{P}_{\theta}(t) = \mathbb{P}(t)+\theta t \dot{\mathbb{P}} (t)+\mathcal{O}(\theta^2)$ and thus we can substitute $ \left.\frac{d P_{i,\theta}(t)}{d\theta} \right|_{\theta=0}= t \dot{P}_i (t)$ in Eq. (\ref{eq:dftheta}).

The condition (\ref{eq:condition1}) does not fully determine the form of the transition rates. To make further progress, 
we choose the following parametrization for the rates
\bea
{K}_{ij,\theta}^{(\alpha)}(t) &=& {K}_{ij}^{(\alpha)}~\exp \left[ \theta \nu_{ij}^{(\alpha)} (t) \right]~,~~~~\nu_{ij}^{(\alpha)} (t)=\left. \frac{\partial}{\partial \theta} \ln {K}_{ij,\theta}^{(\alpha)}(t)\right|_{\theta=0},  \label{eq:biparam} \\
{R}_{ij,\theta}^{(\gamma)}(t) &=& {R}_{ij}^{(\gamma)}~\exp \left[ \theta \mu_{ij}^{(\gamma)} (t)\right]~,~~~~\mu_{ij}^{(\gamma)} (t)=\left. \frac{\partial}{\partial \theta} \ln {R}_{ij,\theta}^{(\gamma)}(t)\right|_{\theta=0} \label{eq:uniparam}.
\eea
 \eref{eq:condition1} is satisfied if we require that
\bea
{K}_{ji}^{(\alpha)}~ \nu_{ji}^{(\alpha)} (t) P_i(t)- {K}_{ij}^{(\alpha)}~ \nu_{ij}^{(\alpha)} (t)P_j(t)={K}_{ji}^{(\alpha)} P_i(t)-{K}_{ij}^{(\alpha)} P_j(t),
\label{eq:currents1}
\eea
for every bidirectional transition. Similarly, for the unidirectional transitions
we demand
\bea
{R}_{ij}^{(\gamma)} ~\mu_{ij}^{(\gamma)} (t)~P_j(t)={R}_{ij}^{(\gamma)} P_j(t).
\label{eq:currents2}
\eea
Equations (\ref{eq:currents1}) and (\ref{eq:currents2}) mean that the partial contribution from each transition conforms with \eref{eq:condition1}.
Note that Eq. (\ref{eq:currents2}) gives $\mu_{ij}^{(\gamma)} (t)=1$, and completely determines the parametrization of the unidirectional transitions.
To determine $ \nu_{ji}^{(\alpha)} (t) $, we substitute \eref{eq:biparam} into
the expression for the Fisher information (\ref{eq:FI-rates}) to obtain
\bea
\mathcal{I}(0)
  &=&\int_0^{\mathcal{T}} dt~ \left[ \sum_{\substack{i,j,\alpha \\ i>j}} \left( P_j(t) {K}_{ij}^{(\alpha)} \left[ \nu_{ij}^{(\alpha)}(t) \right]^2+P_i(t) {K}_{ji}^{(\alpha)} \left[ \nu_{ji}^{(\alpha)} (t)\right]^2 \right)+ \sum_{\substack{i,j ,\gamma \\ i \ne j}} R_{ij}^{(\gamma)} P_{j}(t) \right]~.
  \label{eq:FI-theta0-expand}
\eea
Following \cite{Liu2019}, we connect the terms related to the bidirectional transitions in \eref{eq:FI-theta0-expand} to the entropy production, which is a well known observable in stochastic thermodynamics. Formally, this connection is done by demanding that 
\begin{equation}
 P_j(t) {K}_{ij}^{(\alpha)} \left[ \nu_{ij}^{(\alpha)} (t) \right]^2+P_i(t) K_{ji}^{(\alpha)} \left[ \nu_{ji}^{(\alpha)} (t) \right]^2 = \frac{1}{2} \left[ {K}_{ji}^{(\alpha)} P_i(t) -{K}_{ij}^{(\alpha)} P_j(t) \right] \ln \frac{K_{ji}^{(\alpha)} P_i(t)}{K_{ij}^{(\alpha)} P_j(t)}.
\label{eq:cond2}
\end{equation}
It was shown in  \cite{Liu2019} that Eqs. (\ref{eq:currents1}) and (\ref{eq:cond2}) have a unique time-dependent solution, and thus fully determine the parametrization $\nu_{ij}^{(\alpha)} (t)$. 
One can now substitute \eref{eq:cond2} into 
\eref{eq:FI-theta0-expand} resulting in
\begin{eqnarray}
\mathcal{I}(0)&=&
\int_0^{\mathcal{T}} dt~ \left[ \frac{1}{2} \sum_{\substack{i,j,\alpha \\ i>j}}\left[K_{ji}^{(\alpha)} P_i(t)-K_{ij}^{(\alpha)} P_j(t) \right] \ln \frac{K_{ji}^{(\alpha)} P_i(t)}{K_{ij}^{(\alpha)} P_j(t)} + \sum_{\substack{i,j ,\gamma\\ i \ne j }}  {R}_{ij}^{(\gamma)} P_{j} (t)\right] \nonumber \\
&=& \frac{1}{2}\int_0^{\mathcal{T}} dt~\sigma_\text{rev}(t)+\int_0^{\mathcal{T}} dt~J_\text{uni}(t),
\label{eq:FI-EP}
\end{eqnarray}
where we have introduced 
\bea
\sigma_{\text{rev}}(t)=\sum_{\substack{i,j,\alpha \\ i>j }}\left[K_{ji}^{(\alpha)} P_i(t)-K_{ij}^{(\alpha)} P_j(t) \right] \ln \frac{K_{ji}^{(\alpha)} P_i(t)}{K_{ij}^{(\alpha)} P_j(t)},
\eea
as the entropy production rate due to the bidirectional transitions; and 
\bea
J_{\text{uni}}(t)=\sum_{\substack{i,j ,\gamma \\ i \ne j}}  {R}_{ij}^{(\gamma)} P_{j} (t)~,
\eea
as 
the flux due to the unidirectional transitions. Naturally, $\Sigma_{\text{rev}}=\int_0^\mathcal{T}~dt~\sigma_\text{rev}(t)$ is the total entropy produced due to the bidirectional transitions during the time window $\mathcal{T}$. In contrast, $\Sigma_{\text{uni}}=\int_0^\mathcal{T}~dt~J_\text{uni}(t)$ is the total flux (or activity) of the unidirectional transitions upto time $\mathcal{T}$. Thus, \eref{eq:FI-EP} can be recast in the following way
\bea
\mathcal{I}(0)= \frac{1}{2}~\Sigma_{\text{rev}}(\mathcal{T})+\Sigma_{\text{uni}}(\mathcal{T}).
  \label{eq:FI-EP-2}
\eea
The physical interpretation of both terms in \eref{eq:FI-EP-2} is apparent since it comprises of entropic contributions the bidirectional transitions plus the total flux (or activity) from the unidirectional ones.

What is left is to recast $\left. \frac{\partial F_\theta}{\partial \theta}\right|_{\theta=0}$ [numerator on the RHS of \eref{eq:defgcr}] in term of physical quantities. We now focus on time-independent functionals, assuming that $q_i$, $d_{ij}^{(\alpha)}$, and $c_{ij}^{(\gamma)}$ do not vary during the process.
With the help of $d_{ij}^{(\alpha)}=-d_{ji}^{(\alpha)}$ (the asymmetric property) and a substitution of Eqs. (\ref{eq:currents1}) and (\ref{eq:currents2}) into
\eref{eq:dftheta} we find
\begin{multline}
    \left. \frac{\partial F_\theta (\mathcal{T})}{\partial \theta} \right|_{\theta=0} = \int_0^\mathcal{T} dt \left( \sum_{i=1}^{N_s} q_i t \dot{P}_i (t) + \sum_{\substack{i , j, \alpha \\ i \ne j}} d_{ij}^{(\alpha)} {K}_{ij}^{(\alpha)} t \dot{P}_j (t)+\sum_{\substack{i , j, \gamma \\ i \ne j}} c_{ij}^{(\gamma )} {R}_{ij}^{(\gamma)} t \dot{P}_j \right. \\ +\left. \sum_{ \substack{i , j , \alpha \\ i>j}} d_{ij}^{(\alpha)}  \left[ {K}_{ij}^{(\alpha)} P_j (t)-  K_{ji}^{(\alpha)} P_i (t)\right] + \sum_{\substack{ i , j, \gamma \\ i \ne j}} c_{ij}^{(\gamma)} R_{ij}^{(\gamma)} P_j (t) \right).
    \label{eq:dfdtaftersub}
\end{multline}
We now note that the  rate  of change of $F(t)=\langle \mathcal{F} [\omega] \rangle$ with time can then be written as
\bea
j(t)\equiv \frac{dF}{dt}=\sum_{i=1}^{N_s} P_i(t)q_i+ \sum_{\substack{j , i, \alpha \\ j \ne i}}
  P_j(t)~d_{ij}^{(\alpha)} {K}_{ij}^{(\alpha)} + \sum_{\substack{j , i, \gamma \\ j \ne i}}
  P_j(t)~c_{ij}^{(\gamma)} R_{ij}^{(\gamma)},
\eea
where we have used \eref{eq:dFdt}. We now take the time derivative of the above equation and compare terms with \eref{eq:dfdtaftersub}. After some simplifications, we arrive at the following relation
\bea
\frac{\partial F_{\theta}(\mathcal{T})}{\partial \theta} \bigg|_{\theta=0} = \int_0^\mathcal{T}~dt~ \left[ \frac{d}{dt}\{t j(t)\} -\sum_{i=1}^{N_s}q_i P_i(t) \right]
= \mathcal{T} j(\mathcal{T})-\sum_{i=1}^{N_s}\int_0^\mathcal{T}~dt~q_i P_i(t)~,
\label{eq:numerator-theta0}
\eea
which is the numerator on the RHS of \eref{eq:defgcr}. Plugging back the Fisher information from \eref{eq:FI-EP} and the above relation (\ref{eq:numerator-theta0}) in \eref{eq:defgcr}, the Cram\'er-Rao inequality takes the form
\begin{equation}
\text{Var}_\omega \left[ \mathcal{F}(\omega)  \right] \geq \frac{\left[  \mathcal{T} j(\mathcal{T})-\sum_{i=1}^{N_s}\int_0^\mathcal{T}~dt~q_i P_i(t)  \right]^2}{\int_0^{\mathcal{T}} dt~\left[ \frac{1}{2}\sigma_{\text{rev}}(t)+J_{\text{uni}}(t) \right]}
\label{eq:TUR}.
\end{equation}
Equation (\ref{eq:TUR}) is the central result of this paper. It is a TUR-like relation that holds for models with unidirectional transitions, residence times and for arbitrary initial states. Furthermore, it was
derived for quite general functionals, which are of the form (\ref{eq:functional}). One can find bounds on various physical quantities by
choosing different values for the parameters $q_i$, $c_{ij}^{(\gamma)}$, and $d_{ij}^{(\alpha)}$ of the functional. We apply this relation to several physically interesting examples
in the next section. Before continuing we note that it is possible to generalize the bound also for systems that are externally driven. This can be modeled by considering transition rates that depend on a parameter $\bm{\lambda}$ (not to be confused with the auxiliary parameter $\theta$) that varies with time. The only change in the derivation above is the appearance of an
additional term in $\frac{dj}{dt}$, which results in
\begin{equation}
\text{Var}_\omega \left[ \mathcal{F}(\omega)  \right] \geq \frac{\left[  \mathcal{T} j(\mathcal{T})-\sum_{i=1}^{N_s}\int_0^\mathcal{T}~dt~q_i P_i(t)  -\int_0^\mathcal{T} dt ~t \frac{\partial j}{\partial \bm{\lambda}}\frac{d \bm{\lambda}}{dt}\right]^2}{\int_0^{\mathcal{T}} dt~\left[ \frac{1}{2}\sigma_{\text{rev}}(t)+J_{\text{uni}}(t) \right]}
\label{eq:TURtd}.
\end{equation}
Eq. (\ref{eq:TURtd}) shows some similarities with the TUR recently derived by Koyuk \textit{et al} when there are \textit{only} bidirectional transitions \cite{Koyuk2020}.

\section{Applications}
\label{sec:application}
\noindent
In this section we explore applications of the TUR in \eref{eq:TUR}. We focus on two physical problems that are often described using models with unidirectional transitions, namely stochastic resetting systems and models of enzymatic catalysis. 
The flexibility of \eref{eq:TUR} allows one to apply it to different stochastic quantities, as well as for different types of processes. To highlight this flexibility we apply the TUR to a steady-state system in the stochastic resetting context and to a transient system in the context of enzymatic catalysis.
In each of the examples we obtain a TUR for one
physically relevant quantity that is natural to the problem. Other inequalities can be derived from Eq. (\ref{eq:TUR}) by choosing different functionals.

\subsection{Stochastic resetting systems}
\begin{figure}[t!]
\includegraphics[scale=0.8]{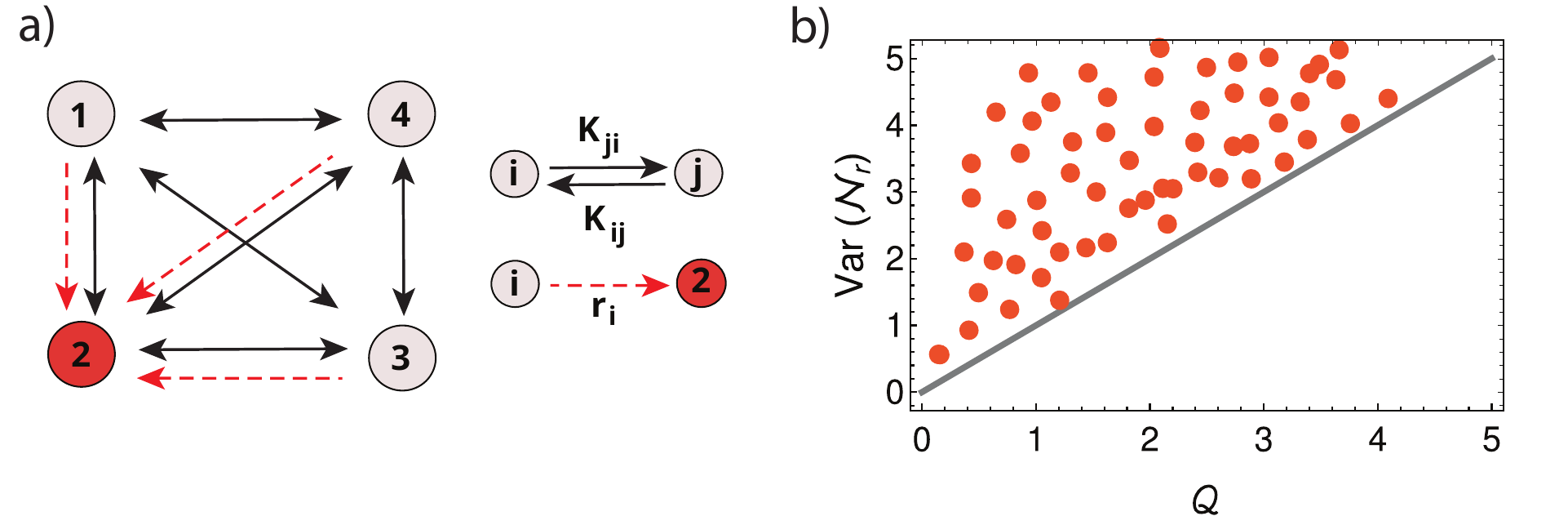}
\caption{Bounds on the fluctuations in the number of resetting transitions. Panel a) A 4-state Markov network with bidirectional (double sided arrows) and unidirectional(single sided arrows) transitions. Bidirectional transitions occur between two states while the unidirectional (resetting) transitions take from state $i=1,3,4$ to the state $i=i_r=2$.  We prepare the system in steady state at time zero and count the total number ($\mathcal{N}_r$) of resetting transitions till an observation time $\mathcal{T}=10$. Panel b) Demonstration of the TUR (\ref{eq:TURresetting}). Here, the variance of $\mathcal{N}_r$ (circle markers in red) is plotted against  $\mathcal{Q}=\frac{\mathcal{T} J_{\text{uni}}^2} {\frac{1}{2} \sigma_\text{rev}+ J_{\text{uni}}}$
for a given realization of the system depicted in panel (a). 
To properly test the TUR, we used random values for the bidirectional and resetting rates, which were taken from a uniform distribution $\mathcal{U}(0.01:10)$. For each such set of rates, we have performed the averaging over $10^6$ stochastic trajectories. As can be seen from the plot, all the results lie above the gray line with slope $1$, in agreement with the TUR of  \eref{eq:TURresetting}.}
\label{fig:resetineq}
\end{figure}

\noindent
Stochastic dynamics with resetting can take place in continuous space, e.g., as in diffusion with resetting  \cite{EM11a}, or alternatively in discrete space by a jump process on a network \cite{Fuchs2016,Pal2017,Busiello2020,network2020}. The TUR derived in the previous section
is relevant for the latter type of dynamics. To model resetting in a jump process on a network one of the states, say $i_r$, is chosen to be the resetting state. Thus, after each resetting event, the system is brought back to that state. Since there are no anti-resetting events the resetting process is modeled by a set of unidirectional transitions that point from any state $i\ne i_r$
to the resetting state $i_r$. We study models in which the resetting process is Markovian, and denote the resetting rates by $r_i=R_{i_r,i}$.
In addition, there are usual bidirectional transitions, with rates $K_{ij}$, between the states and even in absence of resetting the system can move stochastically between the states. 
Thus, the resulting 
stochastic dynamics exhibits a combination of bidirectional transitions, associated with a physical mechanism such as diffusion, and unidirectional transitions describing outside intervention that resets 
the system. An example for such a system is given in  Fig. \ref{fig:resetineq}a.
In this model one can make 
bidirectional transitions (or ``diffuse'') among four states. In addition, the system also undergoes random resetting events that bring it back to state $i_r=2$.

Let us consider such a model and record many histories with the same duration $\mathcal{T}$. To simplify the considerations we assume that the resetting process is autonomous, with time independent rates. We also assume that the system is in steady-state, and denote its probability distribution by $\pi_i$.
A natural quantity to study is the number of resetting events in a realization
\begin{equation}
\mathcal{N}_r [\omega] \equiv \sum_{i\ne i_r} n_{i_r i} [\omega].
\end{equation}
Crucially, this is a functional of the form (\ref{eq:functional}), obtained by substituting $q_i=d_{ij}=0$  and $c_{ij}=1$  for $i=i_r$ and zero
 otherwise.
The rate of resetting events at steady-state $j(\mathcal{T})$ is just the flux to the resetting state
\begin{equation}
j(\mathcal{T}) =\sum_{i \ne i_r} r_i \pi_i = J_{\text{uni}}.
\end{equation}
Thus, the mean number of resetting at steady state is simply given by $\langle \mathcal{N}_r \rangle=\mathcal{T}J_{\text{uni}}$.
Similarly, the steady-state entropy production rate due to the reversible transitions is also time-independent and is given by
\begin{equation}
 \sigma_\text{rev}= \sum_{\substack{i,j,\alpha \\ i>j}}\left[K_{ji}^{(\alpha)}\pi_i-K_{ij}^{(\alpha)} \pi_j \right] \ln \frac{K_{ji}^{(\alpha)}\pi_i}{K_{ij}^{(\alpha)}\pi_j}~,
\end{equation}
and hence 
$ \Sigma_\text{rev}=\mathcal{T} \sigma_\text{rev}$.
The number of resetting events in a process of duration $\mathcal{T}$ therefore satisfies the TUR
\begin{equation}
\frac{  \text{Var} \left[ \mathcal{N}_r \right]}{\mathcal{T} J_{\text{uni}}^2} \ge \frac{1}{\frac{1}{2} \sigma_\text{rev}+ J_{\text{uni}}}.
\label{eq:TURresetting}
\end{equation}
Eq. (\ref{eq:TURresetting}) is obtained from Eq. (\ref{eq:TUR}) by substituting the values of the counting fields and taking into account the fact that the system is at steady state.
We note that when all the resetting rates are equal the TUR can be simplified further since $j(\mathcal{T})=\langle \dot{\mathcal{N}}_r \rangle=r \sum_{i \ne i_r} \pi_i = r \left( 1-\pi_{i_r}\right)$. 

To test the inequality (\ref{eq:TURresetting}), we have considered a 4-state Markov network as shown in \fref{fig:resetineq}a. For given rates we simulated the jump process starting from the steady state distribution. In each simulation we followed the system for a duration $\mathcal{T}$, and counted the number of unidirectional (resetting) jumps that took place during this time. Repeating this process allowed us to calculate the variance of this random variable. The rest of the quantities in  \eref{eq:TURresetting} were calculated from the steady-state distribution. We then repeated the calculations for systems with different values of the rates. These were chosen at random from a 
uniform distribution $\mathcal{U}(0.01:10)$. The results are shown in Fig. \fref{fig:resetineq}b. 
It is clear that the inequality (\ref{eq:TURresetting}) is satisfied by all the examples we tested.

\subsection{Enzyme kinetics}
\label{subsec:enz}
\begin{figure}[h]
\includegraphics[scale=0.5]{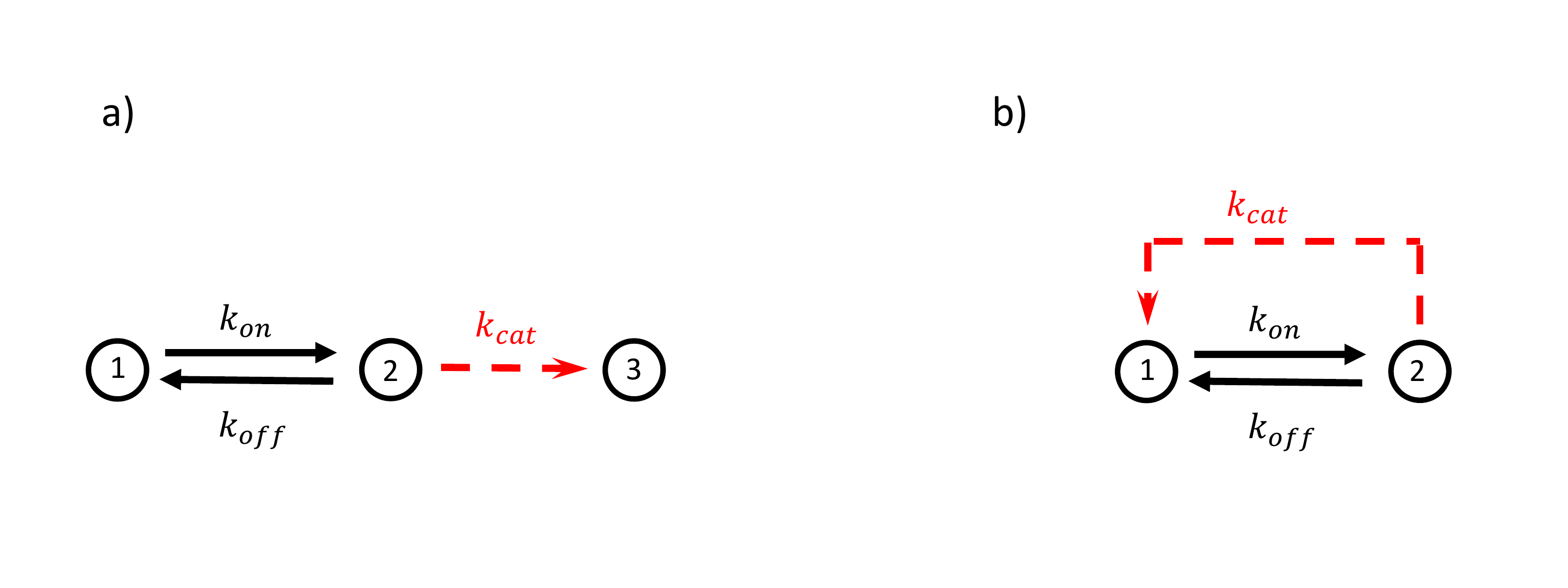}
\caption{A jump process with Michaelis-Menten kinetics. Panel (a) denotes a scheme with transient dynamics that stops once the catalytic step has taken place. In the panel (b) the catalytic step brings the system back to the initial state, thereby allowing to study consecutive catalytic cycles and the enzyme's steady state. \label{fig:mm}}
\end{figure}

\noindent
Enzymatic dynamics can be modeled as Markovian jump processes \cite{enzyme2,enzyme3,enzyme4}. Moreover, such models often include unidirectional transitions.
Fig. \ref{fig:mm} depicts the canonical example of Michaelis-Menten kinetics. According to this model, a substrate molecule binds to the enzyme with a rate $k_{on}$. Once bound to the enzyme the substrate molecule
can either dissociate with rate $k_{off}$, or undergo catalysis 
to form products with rate $k_{cat}$.
The kinetic scheme in Fig. \ref{fig:mm}a can be used to study the dynamics of a single catalytic cycle (essentially a first passage problem that is also conditioned on a catalysis event that occurred at time $0^-$). The kinetic scheme depicted in Fig. \ref{fig:mm}b is obtained by returning the enzyme to its initial state after each catalytic event. It allows one to study the
steady state of the enzyme. This makes the kinetic scheme of Fig. \ref{fig:mm}b very similar to the resetting systems studied above. In particular, a random variable which counts the number of completed cycles in a finite time would satisfy a TUR akin to Eq. (\ref{eq:TURresetting}). To highlight different aspects of the method we instead focus on deriving a TUR for the transient dynamics of the scheme depicted in Fig. \ref{fig:mm}a.

The unidirectional transitions in such models should be understood as approximations, or idealizations of the real reaction schemes in certain limits. They are used either because the reverse transition is 
so rare it is never observed, or if an experiment is stopped once a transition is observed for the first time. The neglected or ignored reverse transition is needed if one wishes
to quantify the entropy production of that step in the cycle. However, the popularity of models with unidirectional transitions means that it will be very useful to be able to apply
concepts such as TURs for their dynamics. Utilizing the theoretical framework developed in Sec. \ref{sec:derivation}, we will derive a TUR for the kinetic scheme depicted in Fig. \ref{fig:mm}a. We then use this simple model to examine the freedom of viewing a bidirectional transition - here the $1 \Longleftrightarrow 2$ transition - as  a pair of unidirectional   transitions. We show that this results in an additional bound and check to see which one is tighter.

\subsubsection{TUR for a Michaelis-Menten model}

The Michaelis-Menten scheme, and many other models of enzymes, are characterized by $n-1$ states that are connected by bidirectional transitions,
and one absorbing state, $n$. The system reaches the absorbing state when the enzymatic cycle is over. The transitions to that state are all assumed to
be unidirectional. We note in passing that models with a more complex structure of unidirectional transitions  can be treated using this general formalism as well. In particular, here we will be interested in the following functional
%Let us consider the following functional
\begin{equation}
\mathcal{C}[\omega] = \sum_{i=1}^{n-1} n_{n,i} [\omega]=n_{3,2} [\omega],
\end{equation}
where the last equality gives the expression of the functional for the Michaelis-Menten scheme depicted in \fref{fig:mm}a. This functional counts the number of irreversible transitions, and is therefore similar to the one studied in the previous subsection. However, in the context of the enzymatic model 
studied here, it acts as a random variable which tells us if the catalytic cycle is complete or not. Thus, 
$\mathcal{C}(t)$ is an indicator function that gets the value $1$ if the enzyme completed its cycle before time $t$, otherwise it is zero. Hence, the mean of this observable is given by 
\bea
\langle \mathcal{C}(t) \rangle=\text{Pr}(\text{cycle completion time}<t)=1-S(t),
\eea
where 
\bea
S(t)=\sum_{i=1}^{n-1} P_i (t)~,
\eea
is the survival probability and $P_i(t)$ is the occupation probability at site $i$ at time $t$. Similarly, the variance of $\mathcal{C}$ can easily be calculated to give
\begin{equation}
\text{Var} (\mathcal{C})= S(t) \left[ 1-S(t)\right].
\label{varC}
\end{equation}
%This happens with probability $P_n (t) = 1 - P_{sur} (t)$. ($P_{sur} (t) = \sum_{i=1}^{n-1} P_i (t)$ is the so called survival probability.)
The TUR given by \eref{eq:TUR} can be readily adapted for the Michaelis-Menten model and this observable. The accumulated flux of irreversible transitions up to time $t$ is given by
\bea
\int_0^t dt' J_\text{uni}(t')=1-S(t)~,
\eea
where we have used the fact that $J_\text{uni}(t)=k_{cat} P_2 (t)=\dot{P}_3(t)$ for the example studied here.
 For the Michaelis-Menten model the reversible entropy production is given by
\begin{equation}
\Sigma_\text{rev} (t) = \int_0^t d t^\prime \left[ k_{on} P_1 (t^\prime) - k_{off} P_2 (t^\prime)\right] \ln \frac{ k_{on} P_1 (t^\prime)}{ k_{off} P_2 (t^\prime)}.
\end{equation}
Finally, for this model the current $j(\mathcal{T})$ in the TUR (\ref{eq:TUR}) is the rate of completing the cycle at time $t$, namely $j(t)=\dot{P}_3 (t) = -\dot{S} (t)$.  %$j(t)=\dot{P}_3 (t) = k_{cat} P_2 (t)$. %Finally, for this simple example the variance of $\mathcal{C}$ is easily calculated as
%\begin{equation}
%\text{Var} (\mathcal{C})= P_{sur} (t) \left[ 1-P_{sur}(t)\right].
%\end{equation}
Collecting everything, the TUR for the probability to complete a cycle at time $t$ can be recast as
\begin{equation}
S (t) \left[ 1-S(t)\right] \ge \frac{t^2 \dot{S}^2(t)}{\frac{1}{2} \Sigma_\text{rev}(t) + 1-S(t)}.
\label{eq:TURmm}
\end{equation}

\noindent
The dynamics of the Michaelis-Menten model can be easily solved. Assuming an initial condition of $P_1(0)=1$, $P_2(0)=P_3(0)=0$, one finds
%\begin{equation}
%\mathbb{P}(t) = \begin{pmatrix} 0 \\ 0 \\ 1\end{pmatrix} - \frac{\Lambda_3+k_{on}}{N_2} \begin{pmatrix} -\Lambda_2 - k_{cat} \\ \Lambda_2 \\ k_{cat}\end{pmatrix} e^{\Lambda_2 t}- \frac{\Lambda_2+k_{on}}{N_3} \begin{pmatrix} -\Lambda_3- k_{cat} \\ \Lambda_3 \\ k_{cat}\end{pmatrix} e^{\Lambda_3 t}.
%\end{equation}
\bea
\mathbb{P}(t) = \begin{pmatrix} 0 \\ 0 \\ 1\end{pmatrix} + \frac{\Lambda_3/k_{cat}}{\Lambda_2-\Lambda_3}  \begin{pmatrix} -\Lambda_2 - k_{cat} \\ \Lambda_2 \\ k_{cat}\end{pmatrix} e^{\Lambda_2 t} + \frac{\Lambda_2/k_{cat}}{\Lambda_3-\Lambda_2}  \begin{pmatrix} -\Lambda_3- k_{cat} \\ \Lambda_3 \\ k_{cat}\end{pmatrix} e^{\Lambda_3 t}.
\eea
Here $\Lambda_{2,3}=\frac{-\sigma \mp \Delta}{2}$, with $\sigma = k_{on}+k_{off}+k_{cat}$ and $\Delta=\sqrt{ \sigma^2 - 4 k_{on} k_{cat}}$, are the eigenvalues describing the decay rates of the probability distribution towards the absorbing state. 
%while $N_2= 2 k_{cat} k_{on} + k_{cat} \Lambda_3 + (k_{on}+k_{off}) \Lambda_2$,
%and $N_3= 2 k_{cat} k_{on} + k_{cat} \Lambda_2 + (k_{on}+k_{off}) \Lambda_3$.

\begin{figure}[t]
\includegraphics[scale=0.55]{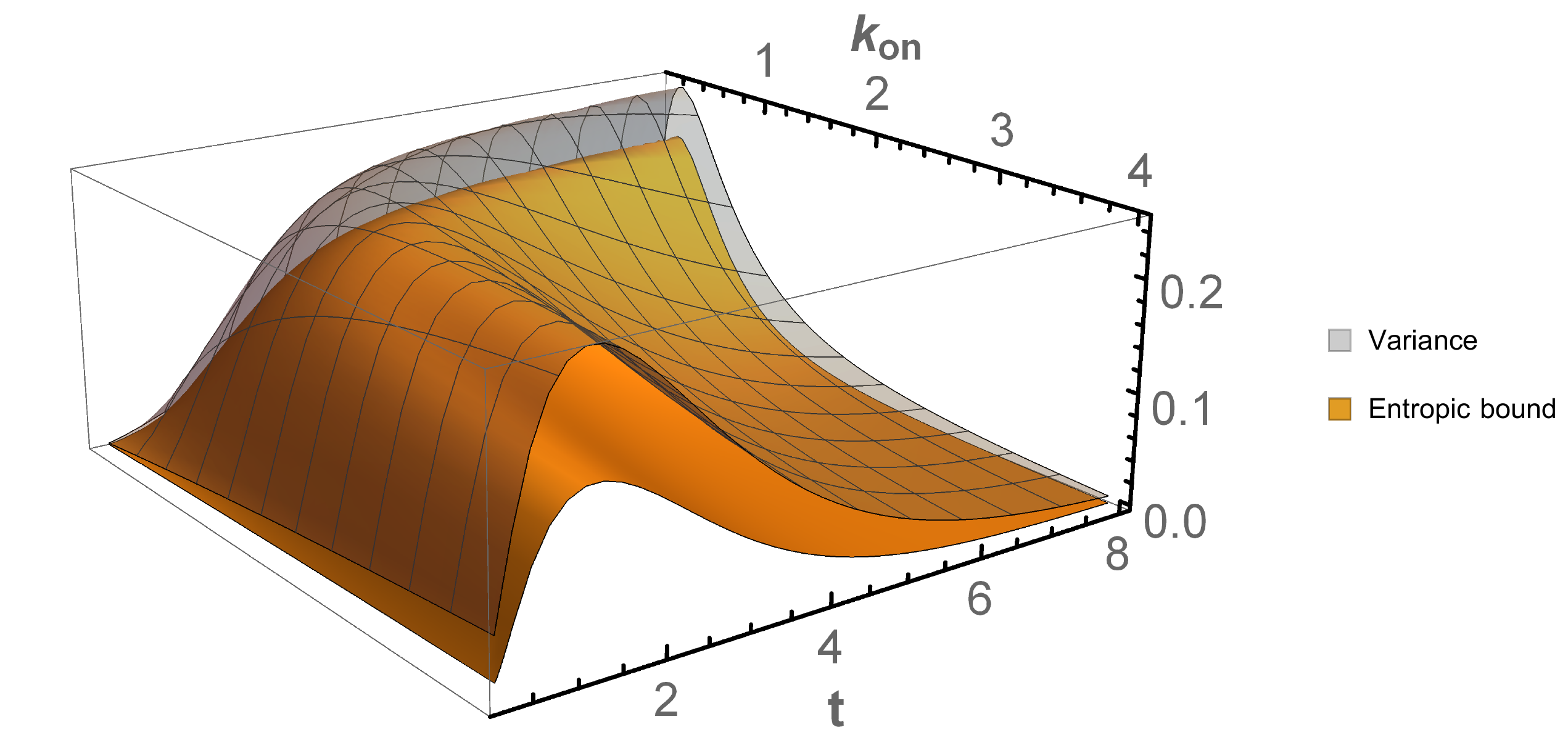}
\caption{The TUR for the Michaelis-Menten scheme in \fref{fig:mm}a. The top meshed surface corresponds to the variance of the random variable $\mathcal{C}$ that indicates whether the cycle is completed by time $t$ [\eref{varC}]. The bottom surface is the lower bound from the right hand side of Eq. (\ref{eq:TURmm}). Here $k_{off}=2$ and $k_{cat}=1. $ } \label{fig:mmineq}
\end{figure}

The two sides of the inequality in \eref{eq:TURmm} are depicted in Fig. \ref{fig:mmineq} for different values of the time, $t$, and binding rate $k_{on}$.
It is clear that the TUR holds for all the parameters included in the figure. Moreover, both surfaces exhibit similar qualitative behavior as the parameters are varied. One should not use the results of Fig. \ref{fig:mmineq} to deduce that the inequality is tight. If one examines the ratio of both sides of Eq. (\ref{eq:TURmm}),
one finds that the ratio is closest to $1$ in the region where the variance is maximal.  The model and observable studied here are quite simple. In particular, the fact that $\mathcal{C}$ can get only two values makes its variance trivially related to its mean. Our results simply demonstrate the validity of the TUR to models of enzymes with transient dynamics. 

\subsubsection{Comparison of entropic and kinetic bounds}

\begin{figure}[t] 
        \includegraphics[scale=0.6]{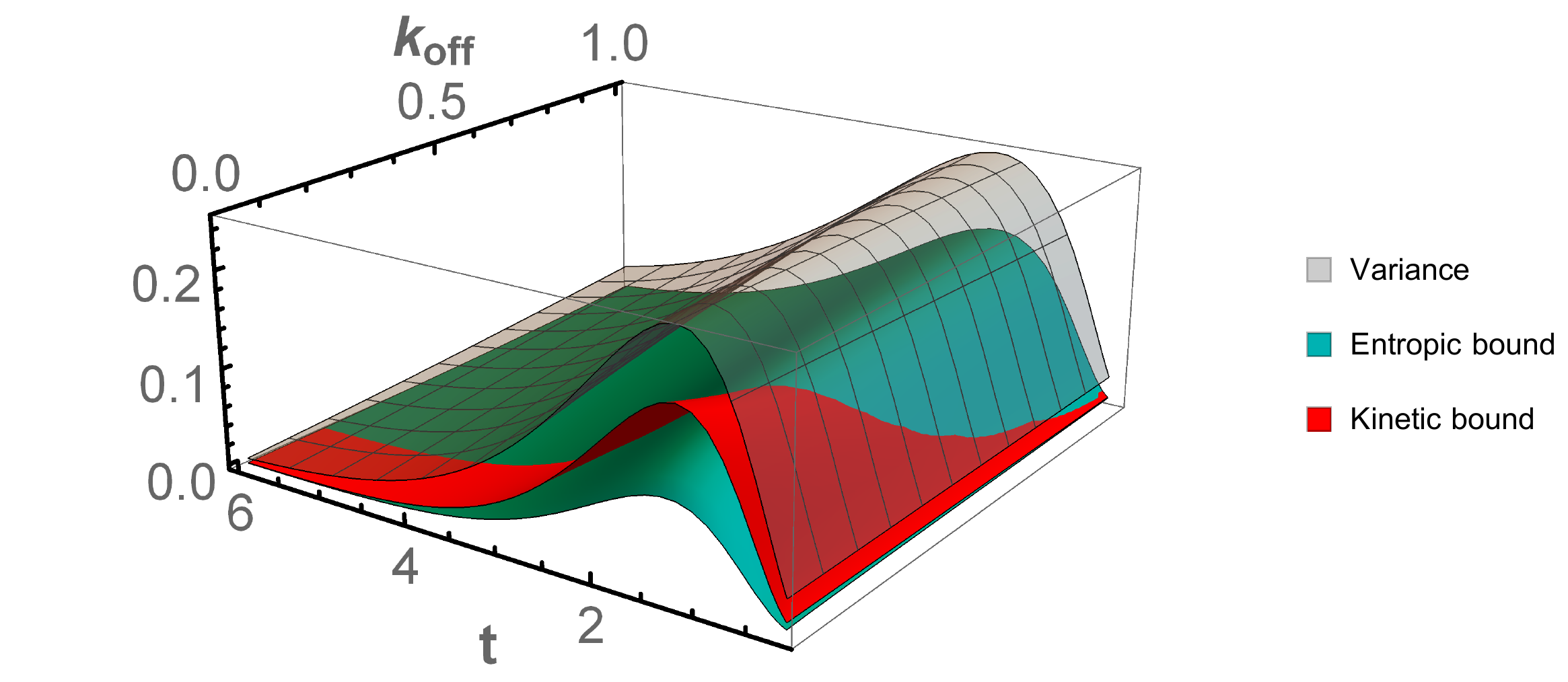} 
 \caption{Comparison of the entropic and kinetic bounds for the Michaelis-Menten scheme. The topmost surface corresponds to the variance of $\mathcal{C}$ from \eref{varC}. %(the left hand side of \eref{eq:TURmm}). 
 The cyan surface is the right hand side of \eref{eq:TURmm} (called here the entropic bound). The red surface 
 corresponds to the right hand side of \eref{eq:TURmm-alt} (or kinetic bound). All surfaces are plotted as a function of $t$ and $k_{off}$, while $k_{on}=2$ and $k_{cat}=1$ are kept fixed.}
 \label{fig:twobounds}
\end{figure}

\begin{figure}[h]
\includegraphics[width=16cm,height=5.5cm]{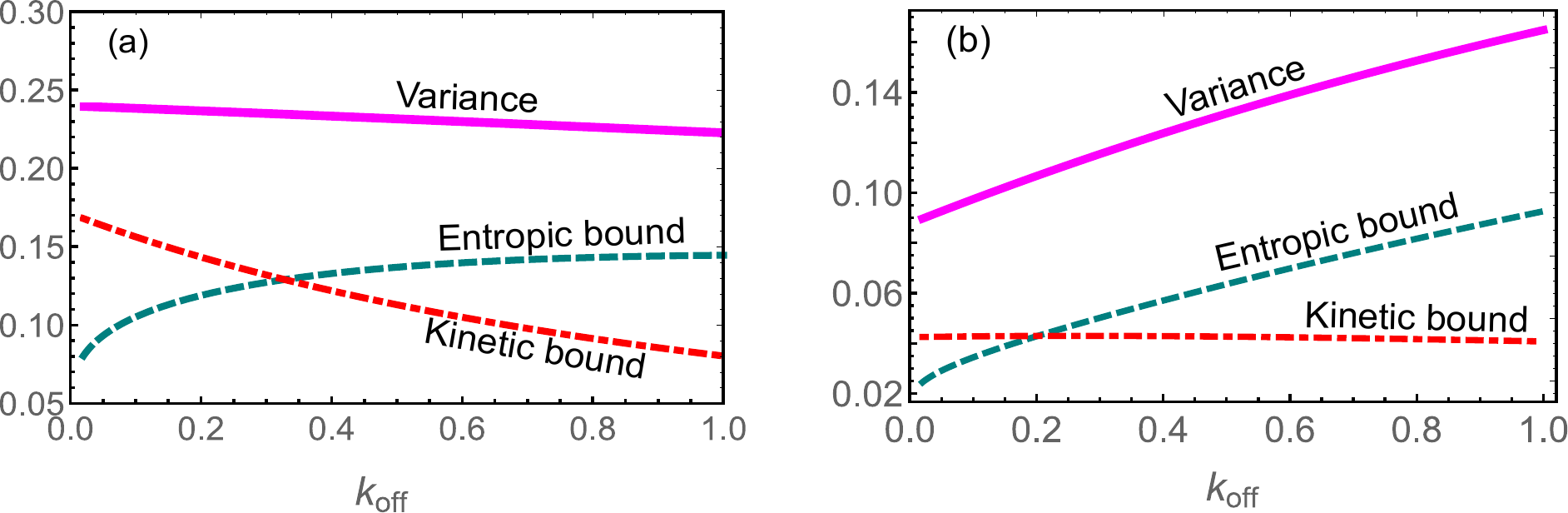}
\caption{One dimensional representation of the results in Fig. (\ref{fig:twobounds}) obtained for fixed $t$. Panel (a) and panel (b) show the cross sections for 
$t=1$ and $t=3$ respectively. Here, we set $k_{on}=2$ and $k_{cat}=1$.}
\label{bounds-limits}
\end{figure}

The simplicity of the Michaelis-Menten model can be used to illuminate a property of the derivation
of the TUR. Namely, one \textit{can choose to treat any bidirectional transition as a pair 
of unidirectional} ones. If we apply this to the $1 \Longleftrightarrow 2$ transitions
in Fig. \ref{fig:mm}a we find an additional inequality
%\sout{$$ S (t) \left[ 1-S(t)\right] \ge \frac{t^2 k_{cat}^2 P_2^2 (t)}{\int_0^t dt' \left[ k_{on}P_1(t')+k_{off} P_2(t') \right] + 1-S(t)}. $$}
\bea
S (t) \left[ 1-S(t)\right] \ge \frac{t^2 \dot{S}^2 (t)}{\Sigma_\text{uni} (t)+ 1-S(t)},
\label{eq:TURmm-alt}
\eea
where 
\bea
\Sigma_\text{uni} (t)=\int_0^t dt' \left[ k_{on}P_1(t')+k_{off} P_2(t') \right].
\eea
Crucially both \eref{eq:TURmm} and \eref{eq:TURmm-alt} are valid inequalities. Since both hold, one should rather ask which one is tighter and therefore more informative. Intuitively, one expects that this depends on specific details of the model, and in particular, how close are the $1$ to $2$ transition to being approximately unidirectional, for instance when the rate $k_{off}$ becomes small.
Figure (\ref{fig:twobounds}) shows a comparison of the ``entropic'' bound from 
\eref{eq:TURmm} and the kinetic bound from
\eref{eq:TURmm-alt}. The upper, meshed surface, is the variance $S (t) \left[ 1-S(t)\right]$ which is plotted as a function of $k_{off}$ and $t$. The cyan surface is the right hand side of \eref{eq:TURmm} whereas the red surface corresponds to the right hand side of \eref{eq:TURmm-alt}.
Figure (\ref{bounds-limits}) depicts two one dimensional cross sections of the surfaces, one at $t=1$ and the other made at $t=3$.

The results in Figs. (\ref{fig:twobounds}) and (\ref{bounds-limits}) show that the tighter bound depends on the values of model parameters. At large values of $k_{off}$ the entropic bound is tighter, and in this case a measurement (or calculation) of the variance $\text{Var} (\mathcal{C})$ will give a useful limitation of the entropy production and a less restrictive one for the integrated fluxes of the $1 \Longleftrightarrow 2$ transitions. In contrast, at small values of $k_{off}$ the more restrictive bound is the kinetic one. This behavior can be understood qualitatively by realizing that the entropy production associated with a $1$ to $2$ transition blows up when $k_{off} \rightarrow 0$. One can apply the same ideas to the  unidirectional $2 \rightarrow 3$  transition. The TURs in \eref{eq:TURmm} and \eref{eq:TURmm-alt} can be viewed as if they were obtained by considering a model with bidirectional transitions, taking the limit of vanishing $3$  $\to 2$ rate, and using the (clearly tighter) kinetic bound for the catalysis step of the cycle.

\section{Discussion and Concluding perspective}
\label{sec:conc}

In this paper we have derived a thermodynamic uncertainty relation that can be applied to models with unidirectional transitions. Such models are used to study a variety of physically relevant processes, including  stochastic resetting systems and enzymatic catalysis which were given here as illustrative examples. Interestingly, the TUR turns out to depend on the entropy production of bidirectional transitions and on the total activity (or flux) of the unidirectional transitions.

The derivation of our main result, Eq. (\ref{eq:TUR}), is based on the Cram\'er-Rao inequality. The derivation is an extension of the one given 
by Liu {\em et. al.} \cite{Liu2019} for \textit{bidirectional transitions} to systems with \textit{unidirectional transitions}. Since the derivation is not based on large deviation theory there is no need to assume that the system is in steady-state. Thus, beyond the ability to describe models with unidirectional transitions, the TUR obtained here can also be applied to processes that may not be at steady-state, such as a single cycle of an enzyme. This gives the freedom to examine the possible role of different initial conditions. 
An interesting time-dependent TUR was recently derived by Koyuk and Seifert \cite{Koyuk2020}. However, we note that their TUR is valid for systems that has \textit{only} bidirectional transitions, and is based on a different mathematical approach.

Unidirectional transitions are often regarded as simplifications, or idealizations, of the real world, since the principle of microreversibility states that if a transition $i \rightarrow j$ is possible, then so is its $j \rightarrow i$ counterpart. One possible exception to this rule is resetting, which is viewed as something that is done by an external agent. One of the problems of models with unidirectional transitions is that the entropy production of those transitions is not well defined. Should this affect the usefulness of the TUR (\ref{eq:TUR})? In fact, the derivation presented above helps to clarify some of the
aspects of the approximation in which one describes a transition as being unidirectional, as explained below.

The derivation of the TUR had considerable freedom. As discussed in Sec. \ref{subsec:enz}, a pair of transitions $K_{ij}$ and 
$K_{ji}$ could be
treated as a bidirectional transition, or as a pair of unidirectional transitions.
This is in fact a \textit{general feature} of the derivation, and is \textit{not restricted} to the Michaelis-Menten model or to a specific transition. Both choices result in different, but valid, inequalities. The difference appears in
the denominator on the right hand side of Eq. (\ref{eq:TUR}).
If the $ij$ transition is treated as bidirectional, the denominator includes a term that expresses the entropy production due to this transition namely
\begin{equation*}
  \Sigma_\text{rev}^{(ij)}=  \int dt \left[ K_{ij} P_j (t) - K_{ji} P_i (t) \right] \ln \left[ \frac{K_{ij} P_j (t)}{K_{ji} P_i (t)}\right].
\end{equation*}
On the other hand, if one chooses a parametrization that treats the $ij$ transitions as two unidirectional transitions, one finds an inequality in which the term above is replaced with 
\begin{equation*}
 \Sigma_\text{uni}^{(ij)}=   \int dt \left[ K_{ij} P_j (t) + K_{ji} P_i (t) \right].
\end{equation*}
It is important to note that both the terms are positive, and they are added to positive contributions from other transitions.

TUR like inequalities can be used to obtain bounds on system structure and properties from experimentally accessible fluctuations of observables. The discussion above points out that many such inequalities are valid, and that the entropy production is not the only relevant quantity. Which inequality should one use? The more informative bound is the one that is tighter. Luckily, finding the tightest inequality can be done by considering each transition separately. We believe that the choice depends on the process one wishes to study. If a pair of transitions $i \rightarrow j$ and $j \rightarrow i$ are close to being detailed balanced during the process, one should use the TUR with $\Sigma_\text{rev}^{(ij)}$, and obtain a bound involving entropy production associated with this transition. If instead one of the transitions is very unlikely, the flux related term $\Sigma_\text{uni}^{(ij)}$ is smaller, and thus the more informative bound will include this term instead of the entropy production, since it is very large in such cases.  This was made clear in the enzymatic catalysis example 
studied in Sec. \ref{subsec:enz}. Following this argument, it is helpful to view unidirectional transitions as a limit in which the rate of a transition goes to zero. In that case the entropy production diverges, and a bound that is based on the entropy production is therefore trivial and non informative as it simply states that the variance is positive. Our derivation, in fact, shows that one can obtain an alternative, and tighter, bound that involves the net flux of transitions.

One of the possible usage of TURs is to deduce information regarding the topology of transitions and other parameters from measurable observables. It will be interesting to find out how to effectively use the freedom, shown in this paper, to choose different parametrizations that result in different but physically meaningful inequalities for this purpose. This is left for future research.

\begin{acknowledgments}
Arnab Pal gratefully acknowledges support from the Raymond and  Beverly  Sackler  Post-Doctoral  Scholarship and the Ratner Center for Single Molecule Science at  Tel-Aviv University. Shlomi Reuveni acknowledges support from the Azrieli Foundation, from the Raymond and Beverly Sackler Center for Computational Molecular and Materials Science at Tel Aviv University, and from the Israel Science Foundation (grant No. 394/19). Saar Rahav is grateful for support from the the Israel Science
Foundation (Grant No. 1526/15), and from
the U.S.-Israel Binational Science Foundation (Grant
No. 2014405).
\end{acknowledgments}

\appendix

\section{Alternative derivation of Eq. (\ref{eq:dFdt}) }
\label{sec:mean}
\noindent
In this appendix we present an alternative derivation of the expression for the mean value of the functional $\mathcal{F} [\omega]$.
To this end,
let us consider the joint probability  %with at time $t$ and with $\mathcal{F}(t)=f$,
\begin{equation}
Q_i(f,t)=\sum_{\omega}\mathcal{P}[\omega] \chi_i(t) \delta\left[ \mathcal{F}(\omega (0:t))-f  \right],
\end{equation}
to find the system in state $i$, and with $\mathcal{F}(t)=f$, at time $t$.
This joint probability has the marginals
\begin{equation}
Q(f,t)=\sum_{i=1}^{N_s}~Q_i(f,t),
\end{equation}
and 
\begin{equation}
P_i (t) = \int df~Q_i(f,t).
\label{P_i-Q_i}
\end{equation}
We wish to write an evolution equation for $Q_i (f,t)$. To do so we identify the various processes that may change $f$ and $i$ in an infinitesimal time step between $t-dt$ and $t$.
For instance, the system will
be in state 
$i$ with $\mathcal{F}[\omega]=f$ at time $t$ if it was at $i$ with $\mathcal{F} [\omega]=f-q_i dt$ at time $t-dt$ and no transition was made in the time interval $dt$. Similarly, if the system was at state $j$ with 
$ \mathcal{F} [\omega] = f-d_{ij}^{(\alpha)}$ at time $t-dt$ it can reach state $i$ with $\mathcal{F} [\omega]=f$ by making the $j \rightarrow i$ transition (via $\alpha$). By including all such incoming and 
outgoing transitions, one arrives at the following evolution equation 
%\begin{multline}
\bea
 \frac{\partial Q_i(f,t)}{ \partial t}=-\frac{\partial Q_i(f,t)}{ \partial f} q_i(t) 
 + \sum_{\substack{j , \alpha \\ j \ne i}} Q_j(f-d_{ij}^{(\alpha)} (t),t) K_{ij}^{(\alpha)} (t) +\sum_{\substack{j, \gamma \\ j \ne i}} Q_j(f-c_{ij}^{(\gamma)} (t),t) R_{ij}^{(\gamma)} (t) -Q_i(f,t)  \lambda_i (t).
 \label{eq:qichange}
\eea
%\end{multline}
Equation (\ref{eq:qichange}) should be supplemented with the initial condition
\begin{equation}
Q_i(f,0)=P_i(0)~ \delta(f).
\end{equation}
We note that the evolution of joint distribution of thermodynamic variables such as work, or entropy production, and the state of the system, is commonly studied in the field (see for instance Refs. \cite{Lebowitz1999,Mazonka1999,Imparato2005,workfluc}). We can now express the mean value of the functional $\mathcal{F}$ as
\begin{equation}
F(t) \equiv \langle \mathcal{F}(\omega) \rangle=\sum_{i=1}^{N_s}~\int df~f ~Q_i (f,t).
\end{equation}
Taking a
time derivative of
both sides, we have
\begin{equation}
\frac{d F}{d t}= \sum_{i=1}^{N_s}~\int df~f ~\frac{\partial Q_i(f,t)}{\partial t}.
\label{eq:derivedfdt}
\end{equation}
Substituting the expression for $\frac{\partial Q_i(f,t)}{\partial t}$ from \eref{eq:qichange} into Eq. (\ref{eq:derivedfdt}) results in 
\begin{align}
\frac{d F}{d t}= \sum_{i=1}^{N_s}~\int d f~f~ \left(- \frac{\partial Q_i(f,t)}{ \partial f} q_i (t) + \sum_{\substack{j , \alpha \\ j \ne i}} Q_j(f-d_{ij}^{(\alpha)}(t),t) K_{ij}^{(\alpha)} (t) +\sum_{\substack{j , \gamma \\ j \ne i}} Q_j(f-c_{ij}^{(\gamma)}(t),t) R_{ij}^{(\gamma)} (t) -Q_i(f,t)  \lambda_i (t)  \right).
\label{eq:df/dt}    
\end{align}

It is not a priory clear what is gained by this substitution, but it turns out that the above expression can be simplified considerably. This is 
done by changing the integration variables of the part that is related to bidirectional transitions. 
After recasting, we have
\bea
\sum_{\substack{j , i, \alpha \\ j \ne i}} \int df ~f ~Q_j \left( f- d_{ij}^{(\alpha)}(t) ,t \right) K_{ij}^{(\alpha)} (t) = \sum_{\substack{j ,i, \alpha \\ j \ne i}} \left [ d_{ij}^{(\alpha)} (t) K_{ij}^{(\alpha)} (t) P_j (t)+\int df' ~ f'~ Q_{j} (f',t) K_{ij}^{(\alpha)} (t)  \right],
\eea
and the terms related to unidirectional transitions can be treated similarly. Substitution of these terms back into 
\eref{eq:df/dt} 
with the use of Eqs. (\ref{eq:defgnd}) and (\ref{eq:defgd}) lead to some cancellations, and we find 
%\textcolor{magenta}{and noting that \textcolor{red}{***Please verify this equation, I think it helps explain a nontrivial transition.***}
%\begin{equation}
%\sum_{i=1}^{N_s}~Q_i(f,t)  \lambda_i (t) =\sum_{\substack{j ,i, \alpha \\ j \ne i}}  Q_{j} (f,t) K_{ij}^{(\alpha)} (t)  + \sum_{\substack{j ,i, \gamma \\ j \ne i}}  Q_{j} (f,t) R_{ij}^{(\gamma)} (t),
%\end{equation} gives
\begin{equation}
\frac{d F}{dt} = -  \sum_{i=1}^{N_s}~\int d f~f ~ \frac{\partial Q_i(f,t)}{ \partial f} q_i (t) +\sum_{\substack{j , i, \alpha \\ j \ne i}}  d_{ij}^{(\alpha)} (t) K_{ij}^{(\alpha)} (t) P_j (t) + \sum_{\substack{j , i, \gamma \\ j \ne i}}  c_{ij}^{(\gamma)} (t) R_{ij}^{(\gamma)} (t) P_j (t).
\label{DF/dt-A}
\end{equation}
Next, one employs integration by parts on the first term on
the RHS of the above expression
\begin{equation}
\int df~f \left(- \frac{\partial Q_i(f,t)}{ \partial f} \right) q_i (t) =-f Q_i(f,t) q_i (t) \bigg|_{f=-\infty}^{f=\infty}+\int df~Q_i( f,t)~q_i (t) =q_i (t) P_i(t),
\end{equation}
where in the last line we have used \eref{P_i-Q_i}.
Here we assumed that for any finite time $\lim_{\left| f\right| \rightarrow \infty} f Q_i (f,t) =0$.
Finally, collecting all the terms together in \eref{DF/dt-A}, we arrive at Eq. (\ref{eq:dFdt}).

\end{document}